# The influence of user personality and rating scale features on rating behaviour: an empirical study

ALESSIA ANICETO, University of Turin, Italy

CRISTINA GENA, Dept. of Computer Science, University of Turin, Italy

FABIANA VERNERO, Dept. of Computer Science, University of Turin, Italy

User ratings are widely used in web systems and applications to provide personalized interaction and to help other users make better choices. Previous research has shown that rating scale features and user personality can both influence users' rating behaviour, but relatively little work has been devoted to understanding if the effects of rating scale features may vary depending on users' personality. In this paper, we study the impact of scale granularity and colour on the ratings of individuals with different personalities, represented according to the Big Five model. To this aim, we carried out a user study with 203 participants, in the context of a web-based survey where users were assigned an image rating task. Our results confirm that both colour and granularity can affect user ratings, but their specific effects also depend on user scores for certain personality traits, in particular agreeableness, openness to experience and conscientiousness.

## 1 INTRODUCTION

Most web-based systems and applications allow users to react to the contents they are offered, expressing support, criticism, interest and emotions [5]. User feedback is fundamental in intelligent systems, which can exploit their knowledge about users to provide personalized interaction, for example by recommending potentially interesting content [2] or adapting the user interface. In addition, such feedback is helpful to other users [12], who can take advantage of it to discriminate between the available options, for example in e-commerce and online review platforms [25]. Content personalization and the provisioning of information about the behaviour of other people can be used in persuasion strategies [8], [13] and to support users to make better choices [20].

User feedback is typically expressed through dedicated interface controls such as rating scales, i.e., graphical widgets characterized by specific features, among which are granularity (i.e., the number of options), presence of a neutral point, labelling [4], use of colour [38] and visual metaphor [11].

Previous research in different fields has shown that rating scale features can bias users' rating behaviour, subtly promoting evaluations that are higher or lower than their actual opinion (see, e.g., [23, 37, 39]). According to Gena et al. [11], user ratings are determined by at least three elements: the item being assessed, rating scale features and user personality. The importance of personality in decision making is widely acknowledged, and its impact on rating behaviour has already received much attention (see, e.g., [32]). However, little research has been devoted to understanding if rating scale features have different effects on the behaviour of individuals with different personalities, one notable exception being the work of Manjur et al. [26], who studied the interplay between scale colour, user personality and culture.

In this paper, we aim at expanding the body of knowledge on this topic. We focus on two scale features, granularity, which is acknowledged a particularly significant impact [4], and colour, which was found to elicit somehow contradictory behaviour patterns [26], and try to understand how they impact the ratings of individuals with different personalities. To this aim, we carried out a user study with 203 participants, in the context of a web-based survey where users were assigned an image rating task. Four rating scales with different colours (red-yellow-green, as in [26], and achromatic) and granularities (3, 5 and 9 points) were compared.

## 2 RELATED WORK

Rating scales are studied in different fields, including economics [33], psychology [30, 40] and statistics [9]. The idea that rating scale features such as granularity [23, 30, 40], labelling [9, 37, 39] or the presence of a neutral point [10, 39] can influence users' rating behaviour is well explored in the survey design area, while it only recently emerged in computer science and, in particular, in the field of intelligent interactive systems and recommenders[1] [4, 11]. Historically, in this field there has been more attention on the design of scales [16, 35], including associated user preferences [6, 38] and cognitive workload [34], and on the impact of user personality on ratings [18, 19, 32].

In the following, we will discuss relevant work which studied how granularity, colour and personality can affect users' rating behaviour.

Granularity. Preston and Colman [30] and Weng [40] both considered Likert-type rating scales and found that scales with more response options offer higher reliability. In a study where participants were asked to repeatedly evaluate their level of happiness using scales with different granularities, Lim [23] found that higher granularity causes significantly higher ratings. Cena et al. [4] carried out three user studies in different contexts and domains, and found that granularity is the feature most responsible for the influence scales can have on users' rating behaviour. More specifically, they found that rating scales with a low granularity (e.g., 3 points) push users to provide more extreme ratings, while scales with an unusually high granularity encourage more nuanced, but also higher ratings than scales with an intermediate granularity (e.g., 5 points), thus confirming previous results from [23].

Colour. Studies on the psychological effects of colour perception and, in particular, on its influence on emotions date back to the works of Goethe. More recently, with their review of psychological literature, Elliot and Maier [7] showed that colour impacts people's affect, cognition, and behavior. Among other things, colour is known to influence consumers' behaviour [31] as well as athletes' performances in sport competitions [17].

As far as rating scales are concerned, Tourangeau et al. [37] observed that the fact of assigning different colours to the endpoints of a scale increases their perceived distance and, consequently, induces users to express higher ratings. Similarly, Bonaretti et al. [1] posited that the use of different colour hues can elicit higher levels of emotional intensity, enhancing the positive/negative valence of the scale endpoints. Differently from [37], however, here the authors hypothesize that an anchor contraction effect is triggered, which pushes users to select intermediate ratings. Manjur et al. [26] carried out a user study where they compared participants' behaviour with differently color-coded scales (red-yellow-green and red-yellow-blue), taking into account their personality and culture (individualism vs. collectivism). They found that colour can induce extroverts and collectivists to express higher ratings than they would do with an achromatic scale. On a slightly different note, Mahbub et al. [25] discovered that users prefer visually informative scales, which make use of colour schemes similar to [26] and emojis, over more neutral ones (i.e., standard 5-star scales). They also showed that people give their "true ratings" when they use the scales they prefer.

---

[1] Recommender systems use information on user preferences to generate personalized suggestions to guide users "to interesting or useful objects in a large space of possible options"[2].

User personality. Personality refers to long-lasting characteristics and behaviours which make a person unique [28]. Although several other models were proposed, the so-called Big Five has received the most attention as a comprehensive model of personality traits. The Big Five includes five traits, each of which represents a range between two extremes [29]: Openness to Experience (O), Conscientiousness (C), Extraversion (E), Agreeableness (A), and Neuroticism (N). Classic work in the recommender system area highlighted that broad differences in user personality can impact their ratings: "one optimistic happy user may consistently rate things 4 out of 5 stars that a pessimistic sad user rates 3 out of 5 stars" [32]. More recently, Hu and Pu [18] conducted a pilot user study aimed at understanding the effect of participants' personality on several dimensions of their rating behaviour, such as the number of rated items or the percentage of positive ratings. A subsequent study was carried out with a larger sample to validate their findings [19]: their results show that individuals who score higher on the agreeableness and conscientiousness traits tend to assign more positive ratings. A similar effect was found for neuroticism, although the authors were uncertain about its interpretation. Finally, as previously mentioned, Manjur et al. [26] introduced personality as a factor to better understand the impact of different colour schemes in rating scales.

## 3 METHODOLOGY

With our experiment, we aimed at understanding how users' personality traits and rating scale features, in particular their colour and granularity, can influence users' rating behaviour. We therefore distributed four different surveys where respondents were asked, apart from providing information regarding their personality, to rate two sets of images (abstract backgrounds and landscapes) using a specific rating scale.

We chose a 5-point achromatic Likert-like scale, with numbers and textual labels at the endpoints and midpoint, as our basic scale. In fact, a 5-point granularity is considered optimal to obtain reliable and valid measurements [21, 22] and Likert-like scales are the standard for questionnaires. Thus, we differentiate from related work which concentrated on rating scales used in web platforms and used 5-star scales as their term of comparison [4, 25, 26]. The other three scales were variants of the basic scale, depending either on their colour or granularity: a 5-point coloured scale, using a red-yellow-green set of contrasting colors as in [26], a 3-point achromatic scale and a 9-point achromatic scale (Figure 1).

We chose to have participants rate images of abstract backgrounds and landscapes because we expect them to elicit no strong, extreme emotions, which might surpass the possible impact of personality and rating scale features.

### 3.1 Experimental design

There are three independent variables: the rating scale (between-subjects), with four levels (5-point achromatic, 5-point coloured, 3-point achromatic and 9-point achromatic scale); the object to evaluate (within-subjects), with two levels (abstract backgrounds and landscapes) and participants' personality (between-subjects), defined according to the Big-Five model. The dependent variable is the rating participants assign to each evaluated object.

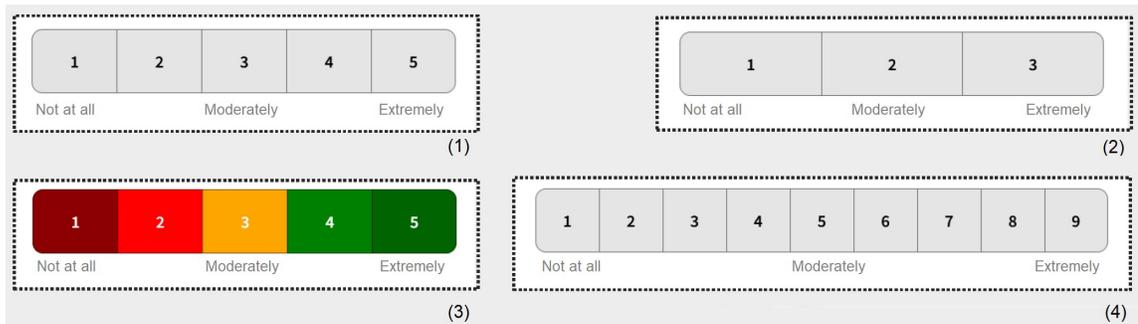

Fig. 1. The four rating scales: 5-point achromatic (1), 5-point coloured (3), 3-point achromatic (2), and 9-point achromatic scale (4).

### 3.2 Hypotheses

We expect that participants' personality has an impact on their rating behaviour and usage of scales which differ because of their granularity and colour. Hence, we formulated the following hypotheses:

H1 Irrespective of participants' personality, rating scales influence their rating behaviour.

H2 Given a certain rating scale, personality influences participants' rating behaviour.

H3 Participants with a certain personality trait behave differently when they use different rating scales.

More specifically, based on previous work, we suppose that the 3-point scale induces participants to assign more extreme ratings (lowest and highest points) [4], while the 9-point scale promotes more nuanced [4], but higher ratings [23], in comparison with the baseline scale (considering that both scales are standardized). In addition, we expect that the 5-point coloured scale pushes participants to express polarized ratings [25, 37], based on their personality; for example, extrovert participants are expected to assign higher ratings [26].

### 3.3 Measures

Participants' personality was measured using the TIPI (Ten Item Personality Measure), a brief measure based on the Big Five model dimensions [15]. TIPI questions ask users to assess how much 10 couples of personality traits apply to them, using a standard 7-point Likert scale ranging from "Disagree strongly - 1" to "Agree strongly - 7". This specific instrument was chosen to keep the test as short as possible and avoid participant fatigue.

### 3.4 Material

Four online surveys were developed using the SurveySparrow[2] platform, each of them integrating one of the rating scales studied in this experiment (see Figure 1). Surveys consisted of different pages, following the steps detailed in the "Procedure" section.

### 3.5 Procedure

The experiment was structured according to the following four steps:

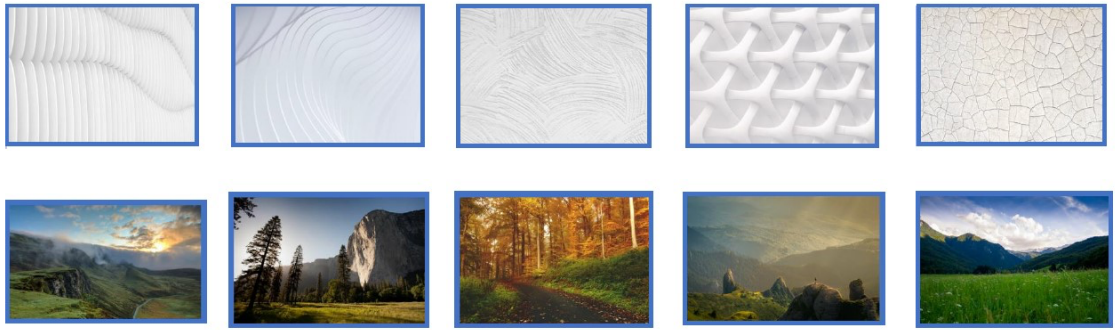

Fig. 2. The ten abstract backgrounds (upper row) and landscape images (lower row) participants were asked to assess.

(1) Welcome. When they accessed one of the four versions of the survey, participants were greeted and introduced to the experiment. In particular, the welcome message contained information on the goal of the study, the proposed scenario for the experimental task (participants were asked to imagine that they were looking for a new background image for their desktop, or for pictures to decorate a waiting room), the task itself (assessing two sets of images), the type of questions they would be asked (personality and demographics), and the expected time to complete the survey (3 minutes). They were also informed that there were no right or wrong answers, and that the survey was completely anonymous. After reading the welcome message, participants were asked to express their informed consent to participate in the study and proceed to the next step.
(2) Image evaluation. Participants were asked to assess 10 images, 5 abstract backgrounds and 5 landscapes, with the rating scale used in their version of the survey (see Figure 2). In particular, for each image they were asked the following question, adapted according to the granularity of the scale they actually had to use: "Considering a 1 (minimum) to 5 (maximum) scale, how much do you like this image?"
(3) Personality test. Participants were asked to answer the ten questions in the TIPI instrument.
(4) Demographics. As a final step, participants were asked to indicate their gender and age.

### 3.6 Participants

---

[2] https://surveysparrow.com/

Participants were recruited through a convenience samplig strategy. Messages containing the links to the surveys were posted on group and personal pages in social networks such as Facebook and Discord, and were shared with friends, relatives and acquaintances through messaging services such as WhatsApp and Telegram.

We collected answers from 203 participants, distributed among the four versions of the survey, more specifically: 55 answers to the survey with the 5-point achromatic scale, 51 to the survey with the 5-point coloured scale, 52 to the survey with the 3-point achromatic scale and 45 to the survey with the 9-point achromatic scale. Most participants (43.8%) fall in the 31-50 age range, followed by those being 51 or more (19.7%), those in the 26-30 age range (18.7%) and, finally, those in the 18-25 age range (17.7%). Most participants identify as female (67%), while 33% of them identify as male.

Table 1. Participants' ratings: descriptive statistics. Normalized values for the 3-point and 9-point scales are included.

|  | 5-pt. achr. | 5-pt. col. | 3-pt. achr. | 9-pt. achr. | 3-pt. achr. norm. | 9-pt. achr. norm. |
|---|---|---|---|---|---|---|
| Minimum | 2.1 | 2 | 1.2 | 3.6 | 2 | 1.99 |
| Maximum | 4.6 | 4.2 | 2.7 | 8.9 | 4.5 | 4.94 |
| 1st quartile | 2.7 | 3.05 | 1.88 | 5.1 | 3.12 | 2.83 |
| 3rd quartile | 3.65 | 3.7 | 2.3 | 6.5 | 3.83 | 3.61 |
| Average | 3.2 | 3.3 | 2 | 5.8 | 3.33 | 3.22 |
| Variance | 0.42 | 0.28 | 0.13 | 1.51 | 0.20 | 0.84 |
| Average (abstract backgrounds) | 2.5 | 2.5 | 1.6 | 4.5 | 2.66 | 2.49 |
| Average (landscapes) | 3.8 | 4.1 | 2.4 | 7.1 | 4 | 3.94 |

Table 2. Significant results for Sec. 4.1 (Mann-Whitney U Test).

| image type | groups | U | z | p | avg. $1^{st}$ gr. | avg. $2^{nd}$ gr. | eff. size |
|---|---|---|---|---|---|---|---|
| landscapes | 5-pt achr. (n = 55) vs 5-pt col. (n = 51) | 1087.5 | -1.98863 | < .05 | 3.82 | 4.12 | 0.19 |
| all | 5-pt achr. (n = 55) vs 3-pt achr. (n = 52) | 1094 | -2.09116 | < .05 | 3.20 | 3.41 | 0.20 |

Notice that the recruitment phase was only interrupted when we obtained at least 10 participants having low and 10 participants having high scores for each personality trait, in each survey version[3]. To this aim, we carried out periodic analyses of the partial data, calculating the first and third quartile for each personality trait so as to identify low and high scores, respectively[4].

## 4 RESULTS

As a preliminary step, average ratings were calculated for each participant and image type. In addition, the ratings expressed through the 3-point and 9-point scales were normalized in order to be compared with those from the 5-point scales. More specifically, to map them on a 1-5 scale, the ratings from the 3-point scale were multiplied by 1.666667 and those from the 9-point scale by 0.555555. Descriptive statistics were then calculated for participants' ratings for all four survey versions, and in particular: average (for abstract backgrounds and landscape pictures separately, as well as for both types of images together), variance, minimum and maximum, first and third quartile (see Table 1).

---

[3] This choice was motivated by the fact that the Mann-Whitney test, which we had planned to use for our analyses, has little power with very small samples (see e.g., [24]).
[4] While norms for the TIPI instrument do exist, they only identify the mean values and were calculated with a population different from ours [14]. [5]We chose the Mann-Whitney U Test because it allows to compare differences between two independent groups and does not require that the data are normally distributed. It is often used when there is a small number of individuals [27].

We can notice that landscape images obtained higher average ratings than abstract backgrounds in all the surveys, and that the average rating for abstract backgrounds always lies in the lower half of the rating scale in use, thus revealing a slightly negative opinion.

In the following, we will present the results of our analyses aimed at assessing the plausibility of our hypotheses. To compare participants' rating behaviour in different situations (i.e., different scale or personality), we carried out an inferential statistical analysis applying the non-parametric Mann-Whitney U Test[5], and we also conducted a correlational study using the Spearman correlation coefficient [5] to better understand the relationship between personality scores and ratings.

Table 3. Significant results for Sec. 4.2 (Mann-Whitney U Test).

| scale | image type | groups | U | z | p | avg. $1^{st}$ gr. | avg. $2^{nd}$ gr. | eff.size |
|---|---|---|---|---|---|---|---|---|
| 5-pt. col. | abstract | low A (n = 18) vs. high A (n = 16) | 85 | 2.01844 | < .05 | 2.86 | 2.33 | .35 |
| 5-pt. col. | abstract | low O (n = 10) vs. high O (n = 22) | 55 | 2.21574 | < .05 | 2.64 | 1.98 | .39 |
| 5-pt. col. | landscapes | low O (n = 10) vs. high O (n = 22) | 53.5 | 2.27672 | < .05 | 4.38 | 3.78 | .40 |
| 5-pt. col. | all | low O (n = 10) vs. high O (n = 22) | 42 | 2.74427 | < .05 | 3.51 | 2.88 | .49 |
| 9-pt. achr. | landscapes | low C (n = 11) vs. high C (n = 13) | 35 | -2.08572, | < .05 | 6.46 | 7.76 | .43 |

4.1    Do rating scales influence participants' rating behaviour?

Firstly, we aim at understanding whether rating scales can influence participants' evaluations, independently of their personality (see H1). To answer this question, we separately examined the ratings obtained by: abstract backgrounds, landscape backgrounds, and all images (abstract and landscape backgrounds together), comparing, for each type of image, the ratings expressed through the 5-point achromatic scale (i.e., our basic scale) and, respectively:

- the 5-point coloured scale,
- the 3-point achromatic scale,
- the 9-point achromatic scale.

Thus, we carried out 9 comparisons in total (3 types of images x 3 "non-basic" rating scales). Significant results are reported in Table 2. In particular, we could observe that:

---

[5] The Spearman correlation coefficient is a nonparametric measure which assesses monotonic relationships between two variables.

(1) When using a *5-point achromatic scale* and a *5-point coloured scale*, ratings assigned to landscape backgrounds are significantly different (U = 1087.5, z = -1.98863, p < .05), with a small effect (.19): individuals using a 5-point achromatic scale give lower ratings than individuals using a 5-point coloured scale.

(2) When using a *5-point achromatic scale* and a *3-point achromatic scale*, ratings assigned to all images taken together (U = 1094, z= -2.09116 p < 0.05) are significantly different, with a small effect (.20): individuals using a 5-point scale give lower ratings than individuals using a 3-point scale.

4.2 Does personality influence participants' rating behaviour?

Secondly, we aim at understanding whether participants' personality influences ratings given with a certain rating scale (see H2). To answer this question, we examined the data collected with each rating scale separately and compared the average ratings for abstract backgrounds, landscapes and all images obtained from participants having low and high scores, respectively, for the same personality trait. This means, for example, that we compared the average ratings assigned to abstract backgrounds by participants with low extraversion, on the one hand, and with high extraversion, on the other hand, using the 5-point achromatic scale, and so on. To this aim, we carried out 60 comparisons in total (3 types of images x 5 personality traits x 4 rating scales); see Table 3 for significant results. To further understand the relationship between personality traits and ratings assigned via a certain rating scale, we carried out the same comparisons using Spearman's correlation coefficient; see Table 4 for significant results. In particular, we could observe that:

(1) When using a *5-point coloured scale*, ratings assigned to abstract backgrounds differ based on participants' level of agreeableness (U = 85, z= 2.01844, p < .05), with a medium effect size (.35): individuals with high agreeableness assign lower ratings than individuals with low agreeableness. However, according to our correlational study, there is no monotonic relationship between agreeableness and ratings.

Table 4. Significant results for Sec. 4.2 (Spearman's correlation coefficient).

| scale | image type | trait | r | p |
|---|---|---|---|---|
| 5-point coloured | landscape | openness (O) | -0.327 | < .05 |
| 5-point coloured | all | openness (O) | -0.27778 | < .05 |
| 9-point achromatic scale | landscape | conscientiousness (C) | 0.3446 | < .05 |

Table 5. Significant results for Sec. 4.3 (Mann-Whitney U Test).

| trait | image type | groups | U | z | p | avg. $1^{st}$ gr. | avg. $2^{nd}$ gr. | eff. size |
|---|---|---|---|---|---|---|---|---|
| high O | abstract | 5-pt. acr. (n = 18) vs 5-pt. col. (n = 10) | 38.5 | 2.44526 | < .05 | 2.84 | 1.98 | .46 |
| low O | landscape | 5-pt. acr. (n = 23) vs 5-pt. col. (n = 22) | 152 | -2.28193 | < .05 | 3.83 | 4.38 | .34 |
| low O | all | 5-pt. acr. (n = 23) vs 5-pt. col. (n = 22) | 163 | -2.03217 | < .05 | 3.13 | 3.51 | .30 |

(2) When using a *5-point coloured scale*, ratings assigned to abstract backgrounds (U = 55, z= 2.21574 p < .05), landscapes (U = 53.5, z= 2.27672, p < .05) and all images taken together (U = 42, z= 2.74427, p < .05) differ based on participants' level of openness, with a medium effect size (.39, .40, and .49, respectively): individuals with high openness consistently assign lower ratings than individuals with low openness. In addition, there is a negative monotonic relationship between openness and ratings assigned to landscapes (r = -0.327, p < .05) and to all images taken together (r = -0.27778, p < .05).

(3) When using a *9-point achromatic scale*, ratings assigned to landscapes differ based on participants' level of conscientiousness (U = 35, z = -2.08572, p < .05), with a medium effect size (.43): individuals with high conscientiousness assign higher ratings than individuals with low conscientiousness. Spearman's coefficient indicates a positive monotonic relationship between conscientiousness and ratings assigned to landscapes (r = 0.3446, p < .05).

### 4.3 Is the effect of personality different when participants are using different rating scales?

Thirdly, we aim at understanding whether people with a certain personality trait rate differently on different scales (see H3). To answer this question, we compared the average evaluations expressed by individuals with a certain score ("high" or "low") for a certain personality trait when using the 5-point coloured, 3-point achromatic and 9-point achromatic scale, respectively, with those expressed through the 5-point achromatic scale. The comparison was repeated for the two types of backgrounds, separately and for the aggregate evaluations (abstract backgrounds and landscapes together).

For example, we compared the scores of highly extroverted individuals with the 5-point achromatic scale and with the 5-point coloured scale, for each type of background and overall.

To this aim, we carried out 90 comparisons in total (3 types of images x 5 personality traits x 2 scores ("low", "high") x 3 "non-basic" rating scales). Significant results are reported in Table 5. In particular, we could observe that:

(1) When using a *5-point achromatic scale* and a *5-point coloured scale*, ratings assigned to abstract backgrounds by individuals with high openness are significantly different (U = 38.5, z = 2.44526, p < .05), with a medium effect size (.46): in particular, ratings are higher when a 5-point achromatic scale is used than when a 5-point coloured scale is used.

(2) When using a *5-point achromatic scale* and a *5-point coloured scale*, ratings assigned to landscape backgrounds (U = 152, z = -2.28193, p < .05) and to all images taken together (U = 163, z = -2.03217, p < .05) by individuals with low openness are significantly different, with a medium (.34) and small to medium effect size (.30), respectively):

in particular, ratings are lower when a 5-point achromatic scale is used than when a 5-point coloured scale is used.

## 5 DISCUSSION

Based on our results, we can confirm that there are rating scale features which influence users' rating behaviour (H1): the 5-point coloured and the 3-point achromatic rating scales both push users to assign higher ratings (Section 4.1). We can also conclude that certain personality traits, in particular, agreeableness, openness and conscientiousness, impact users' rating behaviour (H2), even if this effect was observed only with the 5-point coloured and 9-point achromatic scales (Section 4.2). Regarding H3, we observed that participants with a certain personality can actually behave differently when they use different rating scales. According to our results, this happens in particular to individuals with high openness, who express higher ratings when using a 5-point achromatic scale instead of a 5-point coloured scale, and to individuals with low openness, who, on the contrary, express lower ratings when using a 5-point achromatic scale instead of a 5-point coloured scale (Section 4.3).

One general observation is that the effects we report were not always consistent across all types of images. This may be explained by the fact that, in most cases, the impact of users' real preferences towards the object to evaluate was probably stronger than other factors influencing users' rating behaviour, i.e., user personality and rating scale features [11]. Although we purposely chose images which were expected not to evoke strong emotions, participants still assigned different ratings, on average, to abstract backgrounds and landscapes, showing a greater liking for the latter. Significant differences in the average ratings of all images were found only when comparing the usage of the 5-point and 3-point achromatic scales, when considering participants with different levels of openness using a 5-point coloured scale and when comparing the ratings of individuals with low openness using a 5-point achromatic and a 5-point coloured scale, respectively.

Our findings show some coherence with previous related work. Firstly, they confirm the idea that rating scales which make use of colours promote more extreme and, in particular, more positive ratings [25, 37]. However, differently from Mahbub et al. [25], we found that the coloured rating scale did not have any specific effect on highly extroverted individuals, but seemed to positively affect people with low agreeableness and openness, while also having an opposite effect on individuals with low scores on those traits. Secondly, as far as granularity is concerned, our results regarding the (positively) polarizing effect of 3-point scales affirm previous findings [4]. Thirdly, our observation that the scale with the highest granularity (9 points) induces higher ratings than the baseline scale is coherent with Lim's results [23], although we specifically observed this effect only for individuals with high conscientiousness.

On the other hand, our results might appear to be partially in contrast with those of Hu and Pu [19]: while they found that individuals with high agreeableness provide more positive ratings, our data show that this personality trait is associated with significantly lower ratings. However, Hu and Pu used a binary scale where users could choose between opposite categories such as "like" and "dislike", while our results refer to a 5-point unipolar scale -hence, this difference might be ascribed to the scale.

In general, when comparing our results with related work, we must consider that the objects to evaluate and the rating tasks were also different from ours (for example, freely selecting and rating products from an e-commerce website [19], or rating products already in use [25]), which might well have had an impact on the observed effects and can help to explain the main discrepancies we discussed in this section.

## 6   CONCLUSION

This paper presented an experiment aimed at understanding if rating scale features (in particular: colour and granularity) and personality traits can influence user evaluations in an image rating task. Our findings show that user ratings are impacted by both factors, partially confirming and extending previous results from relevant literature.

Designers of systems which make use of rating scales should be aware of these effects. On the one hand, the use of colours and non-standard granularities could be considered as a sort of nudge[6], which can softly guide users towards a preferable choice, in those cases where there actually is one. For example, following a *microsuasion* approach [8], 3-point scales could be adopted in a context, such as a content-sharing platform, where it is more important to show support for other users and for the contents they created than to precisely measure user preferences, because of their capacity to promote higher ratings.

On the other hand, scales with features which could affect users' rating behaviour should be avoided whenever collecting unbiased ratings, which truthfully reflect users' preferences and opinions, is the priority. This is the case, for example, of recommenders and adaptive systems in general [4], since they provide users with personalised suggestions based on the ratings they expressed. Therefore, based on our results, we suggest to avoid using coloured rating scales, in line with Manjur et al.'s [26] design recommendations, as well as unusually low or high granularities such as 3 or 9 points. On the contrary, 5- or 7-point granularities should be preferred [21, 22].

A limit of the here presented study lies in the relatively small number of participants, which is especially relevant in those cases where only subgroups of users with specific personality traits were taken into account for analysis. In addition, regarding the collection of reliable personality data, Big Five bases on self-report methods, thus users need to fill in a questionnaire to be correctly classified. This approach may suffer from the paradox of the active user [3]: users are motivated to get their immediate task done, and they do not like to fill in long forms prior to using a system. A further limit regards the fact that the rating task was decontextualized, and therefore was more remindful of survey completion than interaction with an adaptive system. This aspect should be kept in mind when considering the applicability of our results to different scenarios. Similarly, granularity and colour may have different impacts in a rating task concerning different (possibly, more sensitive) issues. Finally, regarding colour, we focused on a single scheme (red-yellow-green) and cannot therefore exclude that other chromatic choices could produce different effects. Related to this point, we must also bear in mind that certain colour schemes might not be properly perceived by colour-blind individuals and thus cause inconsistent effects.

Future studies should try to address these limitations, for example by contextualizing the rating task in a recommender, online review or social networking system, as well as take into account other rating scale features which have not yet been studied in connection with personality.

---

[6] A nudge can be defined as "any aspect of the choice architecture that alters people's behavior in a predictable way without forbidding any options or significantly changing their economic incentives" [36].